\documentclass[aps,prb,twocolumn]{revtex4} 

\usepackage{amsmath,amsfonts,amssymb,color}
\usepackage{hyperref}

\usepackage[pdftex]{graphicx}

\begin{document}

\title{Electronic coherence and recoherence in pigment protein complexes: The fundamental role of
non-equilibrium vibrational structures}
\author{A.~W. Chin${}^{1,2}$, J. Prior${}^3$, R. Rosenbach${}^1$, F. Caycedo-Soler${}^{1}$,
S.~F. Huelga${}^1$ and M.~B. Plenio${}^1$}
\address{1 Institute of Theoretical Physics, Universit{\"a}t Ulm,
Albert-Einstein-Allee 11, 89069 Ulm, Germany\\
2 Theory of Condensed Matter Group, Cavendish Laboratory, University of Cambridge,  J J Thomson Avenue, Cambridge, CB3 0HE, UK\\
3 Departamento de F{\'i}sica Aplicada, Universidad Polit{\'e}cnica de
Cartagena, Cartagena 30202, Spain}

\begin{abstract}
Recent observations of oscillatory features in the optical response of photosynthetic complexes
have revealed evidence
for surprisingly long-lasting electronic coherences which can coexist with energy transport. These observations have ignited multidisciplinary interest in the possible role of quantum effects in biological systems, including the fundamental - though still unresolved - question of how electronic coherence can survive in biological surroundings. Here we show that in photosynthetic complexes, non-trivial spectral structures in environmental fluctuations can allow for non-equilibrium processes that lead to the spontaneous generation and sustenance of electronic coherence even at physiological temperatures.
Developing advanced new simulation tools to treat these effects, this new insight provides a firm microscopic basis to successfully reproduce the experimentally observed coherence times in the Fenna-Matthews-Olson complex, and thus sets the ground for the future assessment of the role of quantum effects in photosynthetic light harvesting efficiency.
\end{abstract}

\maketitle
\begin{figure}[h]
\vspace*{-0.25cm}
\includegraphics[width=7.25cm]{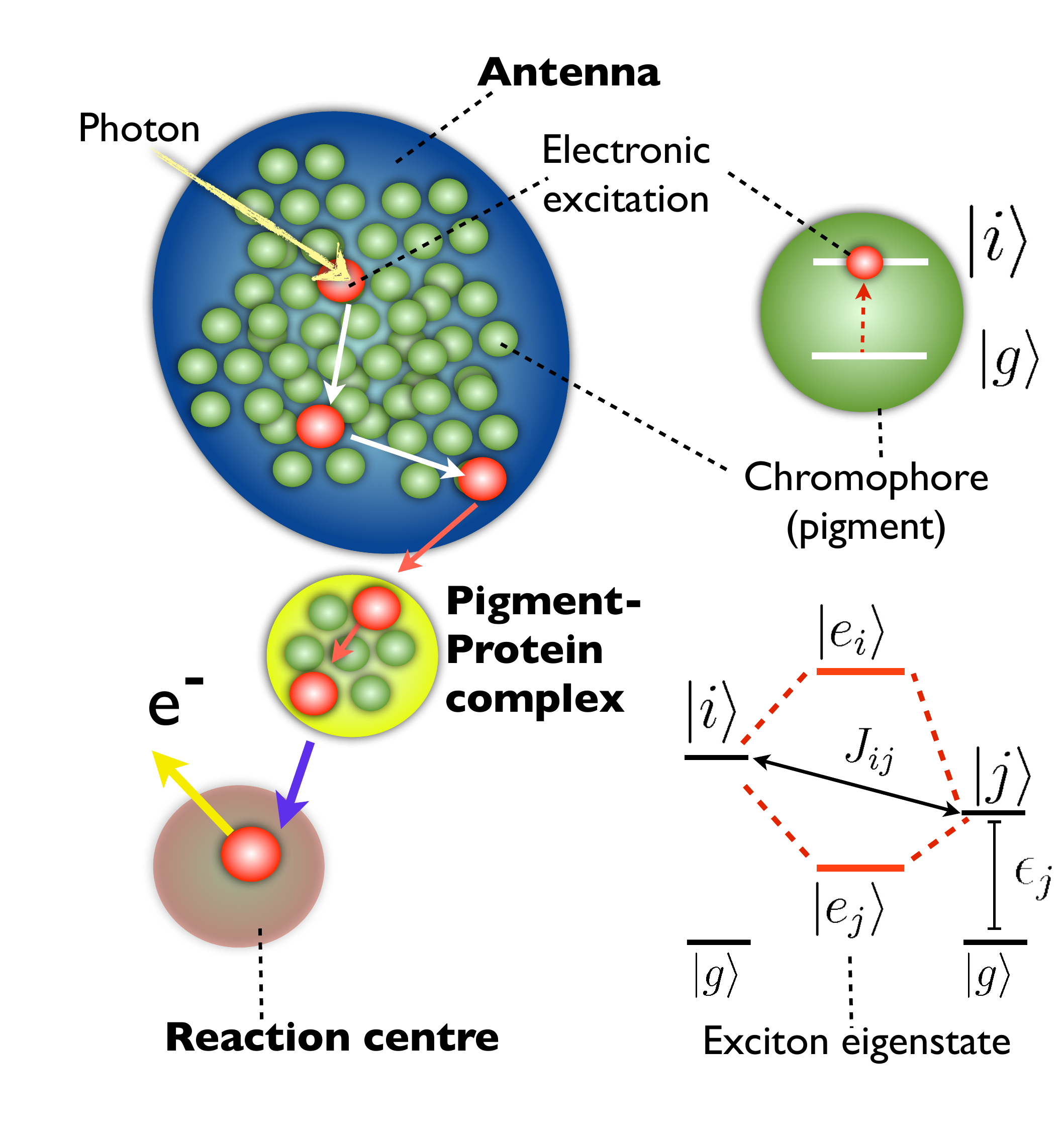}
\vspace*{-0.5cm}
\caption{The generic organisation of the early stages of light-harvesting in natural
photosynthesis. Excitons created in the antenna complexes migrate via dipolar coupling
between chromophores in different pigment-protein complexes and are finally transferred
to a reaction centre where charge separation occurs. Under the low light conditions where
light-harvesting is most efficient, most of the intermediate pigment-protein complexes
transport only single excitons at a time and each complex functions independently. Chromophores
are modeled as having a ground state $|g\rangle$ and optically excited state $|i\rangle$.
Couplings between chromophores leads to the formation of delocalised excitonic eigenstates
$|e_{i}\rangle$, with transitions between these states mediated by environmental fluctuations
}
\label{fig1}
\end{figure}

\emph{Introduction} -- Photosynthesis is a fundamental biological process that provides
the primary source of energy for almost all terrestrial life \cite{blankenship2002molecular}.
In its early stages, ambient photons are absorbed by optically active molecules (pigments)
in an antenna complex, leading to the formation of molecular excited states (excitons).
These then migrate by excitation energy transfer (EET) through pigment-protein complexes (PPCs) to
a reaction centre (RC) where the exciton's energy is used to release an electron - see Fig. \ref{fig1}. Remarkably, these processes
often have a quantum efficiency of almost 100$\%$ \cite{blankenship2002molecular,van2000photosynthetic}, and
uncovering the underlying biological design principles could inspire
important new developments in artificial light harvesting technologies \cite{scholes2011lessons}.
The potential novelty of a biomimetic approach to light-harvesting is underlined by the recent and unexpected
observation of robust, long-lasting oscillatory features in 2-D spectra of PPCs extracted from bacteria, algae
and higher plants. Using ultrafast nonlinear spectroscopy, sustained beating \emph{between} optically
excited states lasting several hundreds of fs at room temperature, and up to nearly $2$ps in
the Fenna-Matthews-Olson (FMO) complex at $77$K, have been observed \cite{gregory2007evidence,panitchayangkoon2010long,calhoun2009quantum,hayes2010dynamics,collini2010coherently}.
These experiments have been interpreted as evidence for \emph{electronic} coherences
between excitons, with lifetimes which are, surprisingly, over an order of magnitude longer than coherences
between electronic ground and excited states \cite{hayes2010dynamics}. Such coherence times are
long enough for EET and excitonic coherence to coexist in these PPCs, conditions under which
a sophisticated interplay of quantum and dissipative processes theoretically optimises transport
efficiency \cite{nat2,plenio08, caruso09,ishizaki2010quantum}.

While many proposals for how quantum effects might enhance biological light-harvesting
have been advanced over the last five years, most of these have used simple, phenomenological
methods to \emph{include} decoherence \cite{nat2,plenio08, caruso09, ishizaki2010quantum,olaya2008efficiency}. In this letter we tackle the more fundamental
problem of elucidating the microscopic mechanisms through which the wide range of electronic
coherence times observed in PPCs actually arise. By including full temperature dependence
into the numerically exact Time Evolving Density with Orthogonal Polynomial Algorithm
(TEDOPA) technique of Refs. \cite{prior10,chinchain10,dmrgbook}, we show that the presence
of resonant structure in the spectral function of PPC chromophores leads to non-equilibrium
quantum dynamics in which picosecond electronic coherence can be driven and supported by quasicoherent interactions between excitons and spectrally sharp vibrational environment
modes \cite{matsuzaki2000energy,Amerongen2000,Ratsep1999,ratsep2007electron}.
At the same time the model also predicts, for the same parameter values, that both the decay
of ground-excited state coherences and the excitation energy transport exhibit the correct
time scales, thus explaining three crucial observations within the same model.
We complement our numerics with an approximate semi-classical model that illustrates the
essential physical mechanisms by which discrete modes, driven far from equilibrium by
exciton injection, may \emph{ spontaneously generate} and sustain oscillatory EET and
electronic coherence against aggressive background decoherence, and which also helps to
isolate the roles of other processes, such as non-Markovian pure dephasing \cite{brumer,kramer}.
Under these conditions, energy transport proceeds in a non-equilibrium fashion, transiently
violating the detailed balance conditions which are often invoked to set fundamental limits
on light harvesting efficiency \cite{scully11}. The conditions that allow for the observation
of this striking behaviour in ensemble experiments are discussed.

\emph{Model} -- Let us consider the following standard Hamiltonian of a PPC consisting of a network
of chromophores which are linearly coupled to environmental fluctuations in their excited states
\cite{nat2,plenio08}. These fluctuations can be due to vibrations in the surrounding protein matrix
and/or intrachromophore vibrational modes.
Defining bosonic annihilation and creation operators
$a_{ik},a_{ik}^{\dagger}$ for the $k-th$ independent oscillator coupled to site ${i}$, and
denoting an optical excitation on site $i$ by the state $|i\rangle$, the Hamiltonian ($\hbar=1$)
for a single electronic excitation can be written as
\begin{equation}
    H = \sum_{i}\left((\epsilon_{i}+X_{i})|i\rangle\langle i| + H^{B}_{i}\right)+\sum_{i\neq j}
    J_{ij} |i\rangle\langle j|,\label{ham}
\end{equation}
where coupling to the vibrational modes of the environment induces local energy fluctuations
via the mode displacement operators $X_{i}=\sum_{k}g_{ik}(a_{ik}+a_{ik}^{\dagger})$ and $H^{B}_{i}=\sum_{k}\omega_{k}a_{ik}^{\dagger}a_{ik}$, with $\omega_{k}$ being the frequencies
of the vibrational fluctuations and $g_{ik}$ their coupling strength to state $|i\rangle$.
The optical excited states $|i\rangle$ have coherent coupling matrix elements (arising
from dipolar coupling) $J_{ij}=J_{ji}$. This coupling  leads to delocalised
\emph{excitonic} eigenstates $|e_{n}\rangle$ with energies $E_{n}$ (we order the states
such that $E_{1}$ is the lowest energy excitation) - see Fig. \ref{fig1}. Finally, we
assume, for simplicity only, identical, independent vibrational environments on each
site characterised by a spectral density  $J(\omega)=\sum_{k}g_{ik}^{2}\delta(\omega-\omega_{k})$. The key findings presented here arise from our consideration of \emph{structured}
spectral functions which contain two contributions; a smooth background describing
fluctuations, likely due to the protein environment, and also couplings to discrete
vibrational modes, which may be of intramolecular origin.

\emph{Physical mechanism of vibration-induced electronic coherence}--Rewriting the Hamiltonian in the excitonic
basis of eigenstates $\{|e_n\rangle\}$, so that $|i\rangle=\sum_{n}C_{n}^{i}|e_{n}\rangle$, it is shown
in the Appendix \ref{hamex} that the exciton-environment interaction develops a transverse (non-adiabatic) term which
couples \emph{different} excitonic states through the bath displacement operators.
Neglecting non-markovian effects, the
background environment leads to incoherent exciton relaxation. However, coupling to quasi-resonant discrete modes allows for the possibility of
\emph{coherent} and reversible inter-exciton transitions via the long-lasting mechanical (coherent) motion
of the mode displacements.

The coupling to the discrete modes is described by the Hamiltonian terms,
\begin{eqnarray}
    H_{I}  &=& \frac{1}{2}\sum_{n,m}(Q_{nm}|e_{n}\rangle\langle e_{m}|+ h.c.),\label{ht}\\
    Q_{nm} &=& \sum_{ik}\sqrt{S_{k}}\omega_{k} C_{n}^{i}C_{m}^{i}(a_{ik}+a_{ik}^{\dagger})\label{r},
\end{eqnarray}
where $\omega_{k}$ are the frequencies of the coherent modes and $S_{k}=(g_k/\omega_k)^2$ are their Huang-Rhys factors
\cite{Amerongen2000}. From Eqs. (\ref{ht}) and (\ref{r}), the initial
(fast) injection of an exciton, {\em either} coherently or incoherently, creates a sudden force on the
modes, which we assume to be initially in a thermal, i.e. incoherent, state. This sudden force initiates
transient oscillations of the modes at approximately their natural frequency $\omega_k$. To first order
these transients can be treated as coherent oscillations - an assumption that will be corroborated later
in our exact numerical treatment - and their back action on the excitons then acts essentially like a
time-dependent field that drives \emph{coherent}, Rabi-like transitions between dissipative exciton states via the Hamiltonian terms
\begin{equation}
    H_{driving}\approx\frac{1}{2} \sum_{n\neq m}(\langle Q_{nm}\rangle (t)|e_{n}\rangle\langle e_{m}|+
    h.c.),
\end{equation}
where $\langle Q_{nm}\rangle (t)\propto \sum_{ik}\sqrt{S_{k}}\omega_{k} C_{n}^{i}C_{m}^{i}\sin(\omega_k t)$ in the above-mentioned approximation of the initial, transient and coherent response of the modes to exciton injection.
Heuristically, this non-equilibrium, laser-like driving essentially generates \emph{new} electronic coherences in the system to
replace those that are continuously damped out by the fluctuations of the smooth background environment. We note that in the language of chemical physics, this phenomenon arises through coherent non-adiabatic coupling (via discrete mode motion) which induces oscillatory crossing of potential surfaces. The actual (coupled) motion of the excitons and modes
is more complex, and may also interact with non-Markovian dynamics of the background, but the physical picture presented here
illustrates a key point: electronic coherence may emerge from transiently exciting robust, \emph{weakly}
dephasing vibrational coherences which are then used to later back-transfer coherence to the excitons, a novel
type of coherence generation and storage (see \cite{PlenioH02,EisertPBH04,HuelgaP07,HartmannDB07,CaiBP08,semiao2010}
for related observations that non-equilibrium systems may generate or maintain quantum entanglement).

We should stress that this (re)generating of electronic coherence is a very
different concept from \emph{protection} of coherences by spatial correlations,
non-Markovianity, or intrinsically weak environmental dephasing
\cite{sarovar2011environmental,ishizaki2010quantum,brumer,kramer}. Crucially, mode-driven coherences will be prominent whenever vibrational modes have
frequencies comparable to exciton energy differences of strongly coupled
chromophores and have dephasing times on picosecond timescales.
This is of particular importance in PPCs, such as the FMO complex, since it
allows regeneration of excitonic coherences whose energy splitting is
commensurate with the frequencies of low energy environment modes.
Many examples of such modes have been observed experimentally in PPCs, such as
the FMO complex \cite{matsuzaki2000energy,Amerongen2000,Ratsep1999,ratsep2007electron}, but their importance for interpreting
experimental observations in multidimensional spectroscopy has only just begun to be appreciated \cite{prior10,caycedo2011nature,christensson2012origin}. This mechanism, based on resonant driving of inter-exciton transitions, will not regenerate or sustain ground to excited state coherence, as the environment does not couple these states. This provides a natural understanding of the distinct coherence time scales in the problem, and we finally note that the tools and ideas we develop here may also be of relevance for photosynthetic charge transfer dynamics, where strong coupling between electronic states due to coherent nuclear motion has been shown to enhance the efficiency of charge separation\cite{novo04}.
\begin{figure}
\includegraphics[width=8cm]{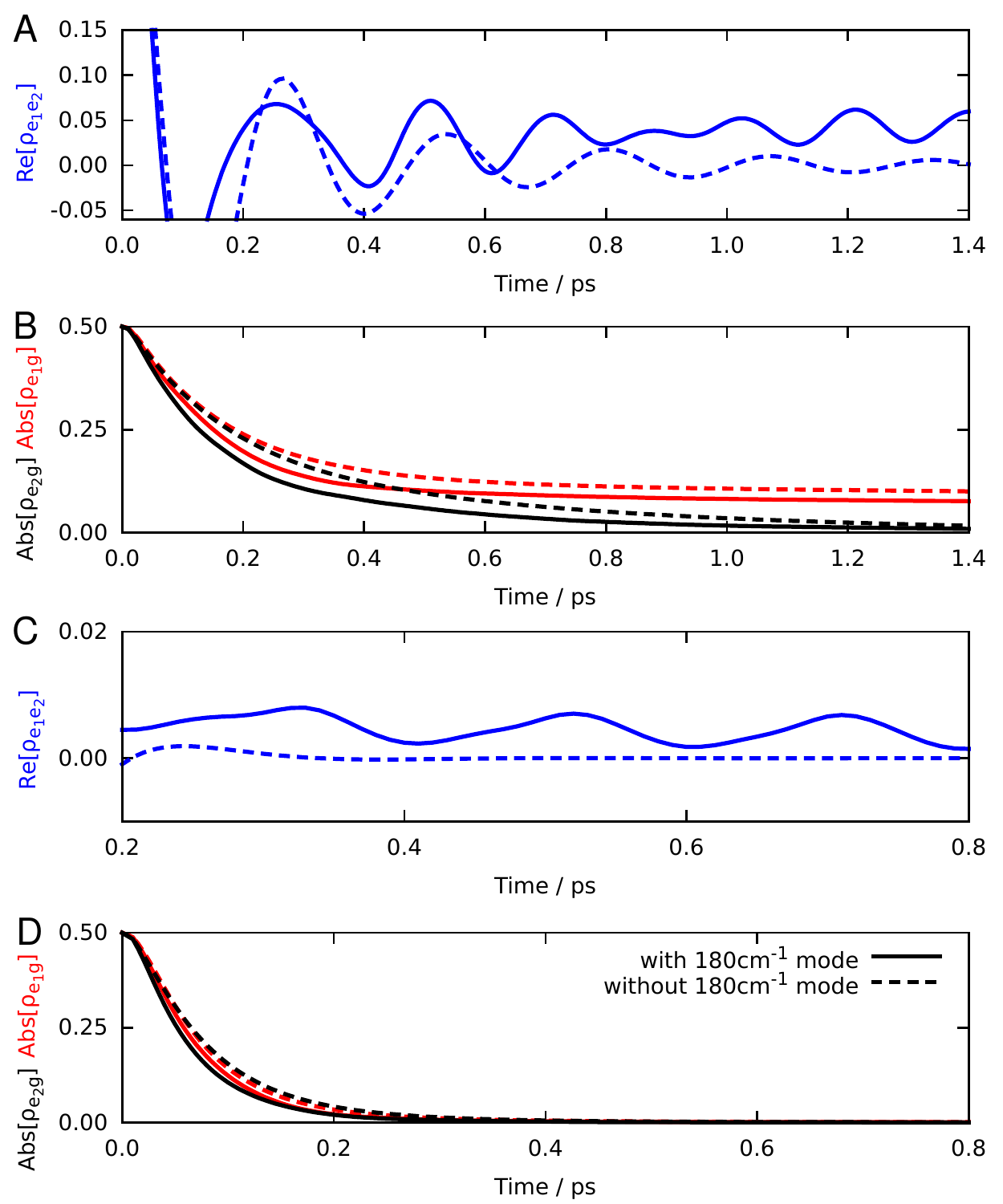}
\caption{In all plots ({\bf A-D}) results including and excluding the resonant $180 \mathrm{cm}^{-1}$ mode are shown as full and dashed lines, respectively. {\bf A\&B}. Semiclassical results at $T=77$K. {\bf A}. Inter-exciton coherence $\mathrm{Re}[\rho_{e_{1}e_{2}} (t)]$ are shown for initial exciton state $2\rho(0)=|e_{1}\rangle\langle e_{1}|+|e_{2}\rangle\langle e_{2}|+|e_{1}\rangle\langle e_{2}|+
|e_{2}\rangle\langle e_{1}|$. {\bf B}. Ground-excited coherences $\mathrm{Abs}[\rho_{e_{1}g} (t)]$,$\mathrm{Abs}[\rho_{e_{2}g} (t)]$ for initial exciton state$2\rho(0)=|e_{i}\rangle\langle e_{i}|+
|g\rangle\langle g|+|e_{i}\rangle\langle g|+|g\rangle\langle e_{i}| (i=1,2)$, respectively. Note that the optical high-frequency oscillations of the $\rho_{e_{i}g}(t)$ coherences have been suppressed by taking the absolute value. Exciton population dynamics following the injection of  an excitation on site 1 of the FMO complex were also computed at this temperature, showing that excitons relax to the lowest energy states localised around sites 3 and 4 after $\approx 3$ ps (not shown). This is inline with the experimental transport times of several picoseconds \cite{adolphs2006proteins}. {\bf C \& D}. Semiclassical results at $T=277$K. The same quantities and initial conditions as {\bf A \& B} are plotted and used in {\bf C \& D}, respectively.  The results at $T=277$K in {\bf C \& D} are plotted on the interval
[0.2,0.8] ps to highlight the long lived coherences in the presence of the resonant
$180 \mathrm{cm}^{-1}$ mode on timescale relevant for experiments at this temperature.}
\label{semis}
\end{figure}

\emph{Fundamental mechanisms supporting long lived coherence: Semiclassical results for the Fenna-Matthews-Olson complex} -- To illustrate the physical mechanism of coherence regeneration through
\emph{coherent} (deterministic) motion of discrete modes in excitonic dynamics, we now present semiclassical simulations obtained by treating the coordinates
and momenta of the discrete vibrations as classical variables. This creates
an effective time-dependent Hamiltonian for the excitons, and we use this in a secular Bloch-Redfield
master equation with time-dependent dephasing rates for the reduced exciton state $\rho(t)$, taking the smooth part of the experimentally-fitted spectral function of Adolphs and Renger (AR) for the
background bath \cite{adolphs2006proteins}. In the following simulations, interactions with a mode of frequency $180 \mathrm{cm}^{-1}$ and $S_{k}=0.22$ were included, and the explicit form of the spectral function is given in Appendix \ref{spectralfunction}. The mode parameters and dephasing rate  ($1 \mathrm{ps}^{-1}$) are taken from hole burning or estimated from fluorescence line narrowing experiments, as is the background spectral function  \cite{adolphs2006proteins,matsuzaki2000energy,Amerongen2000}.
The background bath and modes are initially in thermal equilibrium at temperature $T$, the background reorganisation energy is $\lambda=35$ cm$^{-1}$ and the seven-site
FMO Hamiltonian of \textit{ C. tepidum} was taken from \cite{adolphs2006proteins}.  We denote the matrix
elements $\langle e_{n}|\rho(t)|e_{m}\rangle= \rho_{e_{n}e_{m}}(t)$. A full description of this semiclassical
approach and the importance of time-dependent pure dephasing for super-Ohmic spectral densities is set out in Appendix \ref{semiclassics}.

Figure \ref{semis} {\bf A} shows results at $T=77$K. Motivated by current experimental data \cite{panitchayangkoon2010long}, which focuses on the two lowest exciton states, we present in  Figure \ref{semis} (bottom) the coherences $\mathrm{Re}[\rho_{e_{1}e_{2}} (t)]$ between
exciton states ($|e_{1}\rangle,|e_{2}\rangle$) with a symmetric superposition of these as the initial state to resemble the laboratory condition after excitation with multiple laser pulses. Note that these results are obtained from a full simulation of the FMO complex, and so the influence of all sites on the lowest two excitons are included.
In the presence of the $180 \mathrm{cm}^{-1}$ modes, which are nearly resonant with $E_{1}-E_{2}=150 \mathrm{cm}^{-1}$, after a transient behaviour in the first $250$fs
the coherence exhibits prominent oscillations with an effective coherence time of
$\approx 950$ fs.
This is consistent with the coherence times of beating signals seen in FMO experiments \cite{panitchayangkoon2010long}. In contrast,
for the same background spectral density but neglecting the $180 \mathrm{cm}^{-1}$ mode, oscillations decay
with a coherence time of just $280$ fs after $250$ fs.  A multi-frequency beating pattern in the oscillations is apparent in the presence of resonant modes until about $850$ f. This is caused by the beating and dynamical coupling
between the electronic coherent oscillations caused by the initial laser excitation and those induced later by the discrete mode motion. The vanishing of this pattern at the same rate as the exciton coherence in the background-only simulation provides direct evidence that initial coherences are not protected but \emph{replaced} by mode interactions at later times. Both results also show a fast, time-dependent pure dephasing component at early times due to the background environment, described in detail in Appendix \ref{semiclassics}.

Figure \ref{semis} {\bf B} also shows the absolute value of the coherence between $|e_{1,2}\rangle$ and the optical ground state $|g\rangle$, starting from a symmetric superposition of $|e_{1,2}\rangle$ and $|g\rangle$ which closely resembles the evolution probed through accumulated photon echo experiments \cite{louvwe} and recently developed single molecule femto-second pulse shaping techniques \cite{hildner2010femtosecond}. An additional faster component to the coherence decay is seen for both coherences at early times when the $180 \mathrm{cm}^{-1}$ mode is included, but the dynamics are qualitatively similar to the background-only case. The residual, slowly decaying component of $\rho_{e_{1}g}$ is due to the relatively long  ($\approx 2.4$ ps) lifetime of the lowest energy state at $T=77K$ and the absence of pure dephasing in the long time limit in the AR spectral function in the Markov approximation - see Appendix \ref{nonmarkovsec}. 

Figures \ref{semis} {\bf C} and {\bf D} show corresponding results at $T=277$ K. Coherent oscillations lasting up to at least $600$ fs have been observed in FMO \cite{panitchayangkoon2010long} at this near-ambient temperature. Here we find that the mode interactions lengthen the effective coherence time relative to the background-only simulations even more dramatically, with mode-induced coherence lasting up to $800$ fs, compared with just $200$ fs in the absence of the mode. The $\rho_{e_{1,2}g}$ coherences all decay with a similar, monoexponential time constants in the range $80-100$ fs, which are dominated by the short (and almost equal) lifetimes and transient pure dephasing rates of the exciton state populations at $T=277$K.



\emph{Numerically-exact simulations}-- The essentially Markovian treatment presented so far has served to illustrate the principal physical mechanism responsible for long-lived electronic coherences represents an
approximation whose validity must be assessed. Therefore we now present numerically exact results which include all
possible effects of discrete mode motion, non-adiabatic coupling and fluctuations, as well as the non-Markovian background. For clarity the new physics
are presented for a dimer PPC, an important component of a range of natural PPCs \cite{collini2010coherently,theiss2007pigment,panitchayangkoon2010long}. We again consider the same background spectral
density $J(\omega)$ and $180 \mathrm{cm}^{-1}$ discrete mode of AR \cite{adolphs2006proteins}. A $37 \mathrm{cm}^{-1}$ mode (with $S_{k}=0.1$) which has recently been used to describe features of 2D FMO spectra is also included \cite{caycedo2011nature}, with parameters taken from fluorescence line narrowing experiments \cite{Amerongen2000}.
Electronic parameters used in the simulations are $J_{12}=53.5 \mathrm{cm}^{-1}$, $\epsilon_{1}-\epsilon_{2}
= 130 \mathrm{cm}^{-1}$. This gives two exciton eigenstates with an energy difference of $170 \mathrm{cm}^{-1}$,
which is based on sites $3$ and $4$ of the FMO Hamiltonian \cite{adolphs2006proteins}, and which
are fairly typical values for exciton pairs in PPCs \cite{adolphs2006proteins,gregory2007evidence,panitchayangkoon2010long,calhoun2009quantum}.
The evolution of the \emph{global} system-environment density matrix is then computed
using a new finite-temperature extension of the numerically exact TEDOPA method \cite{prior10,chinchain10,dmrgbook}
- see Appendix \ref{dmrg}. In all simulations the initial state is a product of an exciton state
and a thermal state of the environment (which includes the discrete modes). To show that the main effects arise from the presence of the resonant $180 \mathrm{cm}^{-1}$ mode, simulations
are carried out both with and without this mode, whilst the non-resonant $37 \mathrm{cm}^{-1}$ mode is always retained.
\begin{figure}
\includegraphics[width=8.5cm]{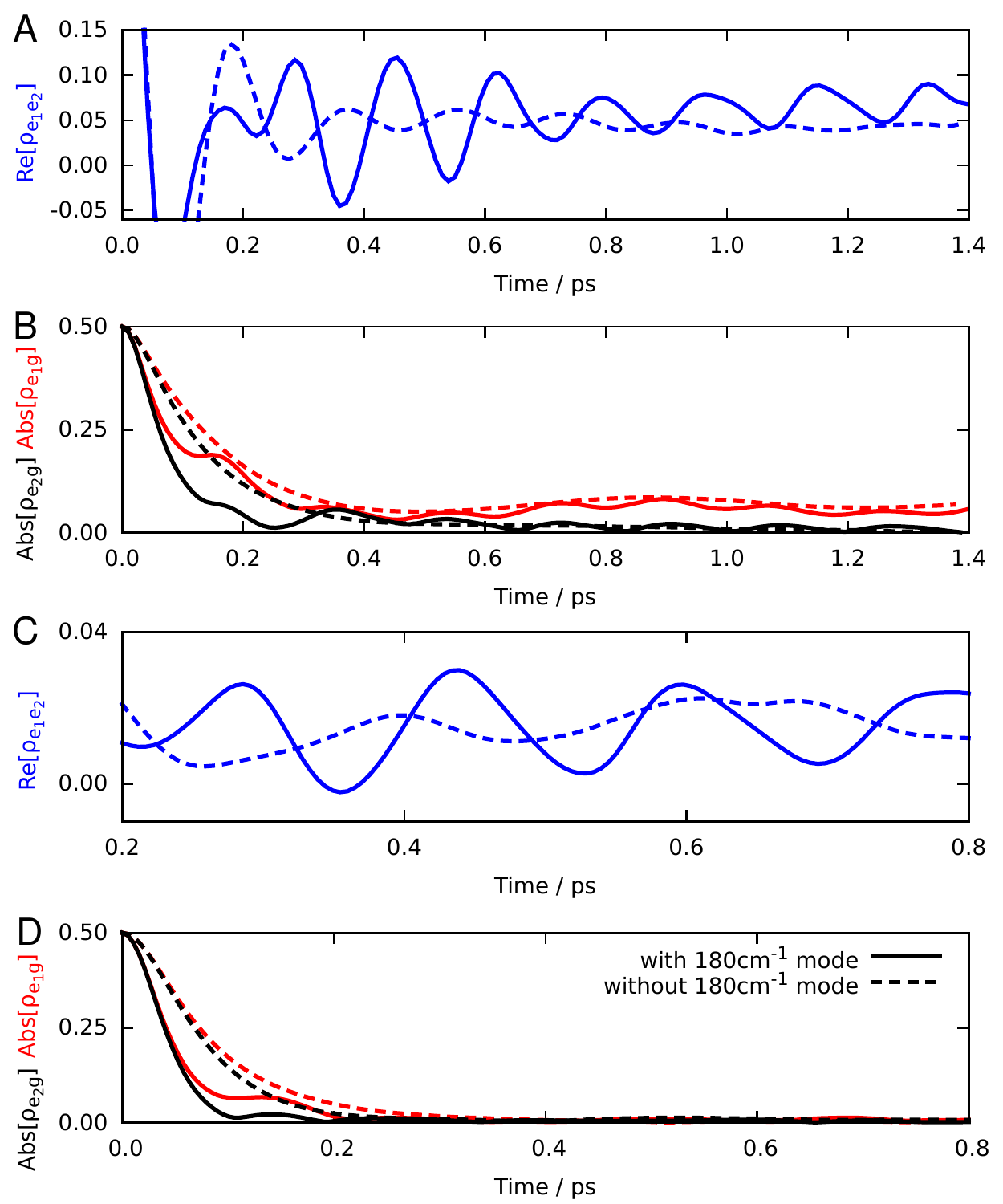}
\caption{ In all plots ({\bf A-D}) results including and excluding the resonant $180 \mathrm{cm}^{-1}$ mode are shown as full and dashed lines, respectively. {\bf A\&B}. TEDOPA results at $T=77$K. {\bf A}. Inter-exciton coherence $\mathrm{Re}[\rho_{e_{1}e_{2}} (t)]$ are shown for initial exciton state $2\rho(0)=|e_{1}\rangle\langle e_{1}|+|e_{2}\rangle\langle e_{2}|+|e_{1}\rangle\langle e_{2}|+
|e_{2}\rangle\langle e_{1}|$. {\bf B}. Ground-excited coherences $\mathrm{Abs}[\rho_{e_{1}g} (t)]$,$\mathrm{Abs}[\rho_{e_{2}g} (t)]$ for initial exciton state$2\rho(0)=|e_{i}\rangle\langle e_{i}|+
|g\rangle\langle g|+|e_{i}\rangle\langle g|+|g\rangle\langle e_{i}| (i=1,2)$, respectively. Note that the optical high-frequency oscillations of the $\rho_{e_{i}g}(t)$ coherences have been suppressed by taking the absolute value. Weak revivals of $\mathrm{Abs}[\rho_{e_{2}g} (t)]$ (on top of faster amplitude modulations) in the interval $[0.3, 0.6]$ ps arise from coherent population transfer (\emph{against} the energy gradient) and match similar features seen in Fig. \ref{population} {\bf A}.    {\bf C \& D}. TEDOPA results at $T=277$K. The same quantities and initial conditions as {\bf A \& B} are plotted and used in {\bf C \& D}, respectively.  The results at $T=277$K in {\bf C \& D} are plotted on the interval
[0.2,0.8] ps to highlight the long lived coherences in the presence of the resonant
$180 \mathrm{cm}^{-1}$ mode on timescale relevant for experiments at this temperature.}
\label{exc1exc2}
\end{figure}

\emph{Inter-exciton and ground-exciton coherences}-- Figure \ref{exc1exc2} {\bf A} shows the evolution of
the electronic coherence $\rho_{e_{1}e_{2}}(t)$ starting from an initially prepared (pure) symmetric
superposition of the exciton states $|e_{1}\rangle$ and $|e_{2}\rangle$ at $T=77$ K. With coupling to the
$180 \mathrm{cm}^{-1}$ mode, the multi-frequency beating
and revival dynamics in the oscillating coherence are again seen, indicating mode-driven coherence. This leads to coherence oscillations with a fast initial dephasing time ($200$ fs) and a residual, longer lasting component arising from mode driving. Similar revival
patterns have been observed in time resolved spectra of FMO \cite{hayes2010dynamics,caycedo2011nature}.
At $T=277$ K, shown in Fig. \ref{exc1exc2} {\bf C}, the resonant mode also greatly enhances coherent oscillations relative to the background-only coherences over the first $800$ fs, which is, again, consistent with experimental results \cite{panitchayangkoon2010long}. The non-resonant mode at $37 \mathrm{cm}^{-1}$ plays no significant role in these inter-exciton dynamics, as expected.

\begin{figure}[t]
\includegraphics[width=8cm]{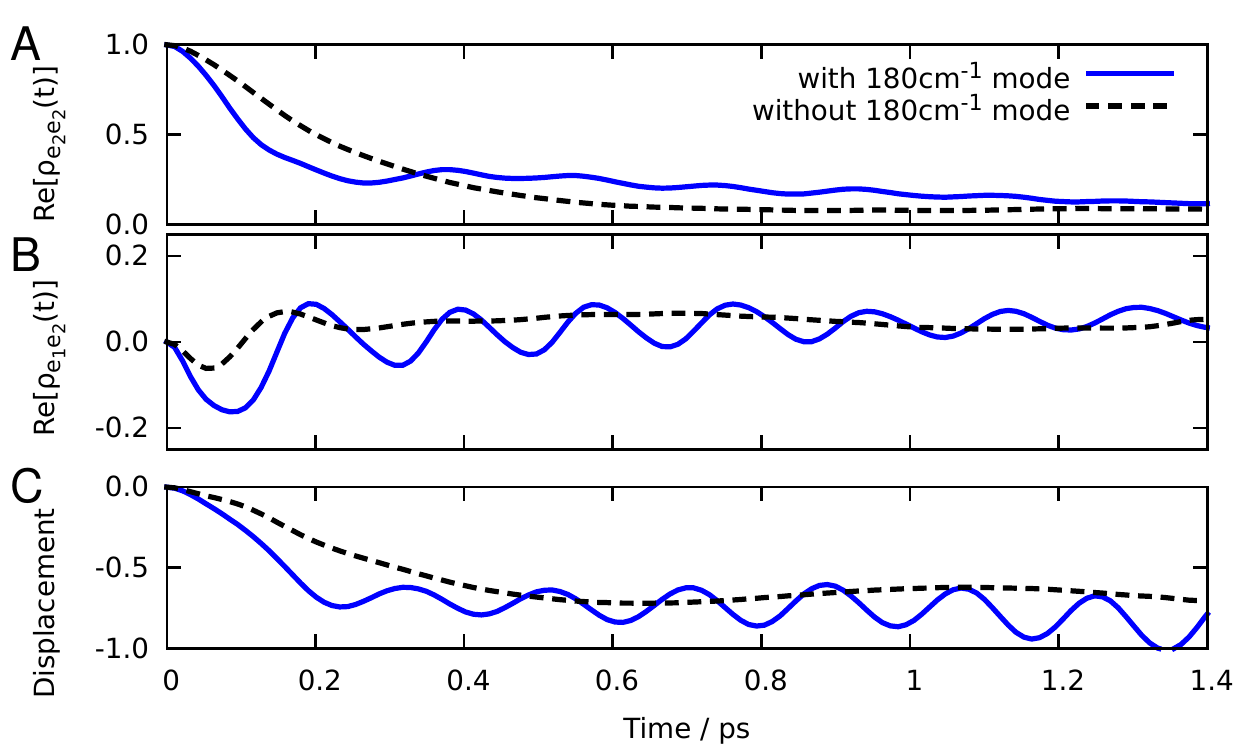}
\includegraphics[width=8cm]{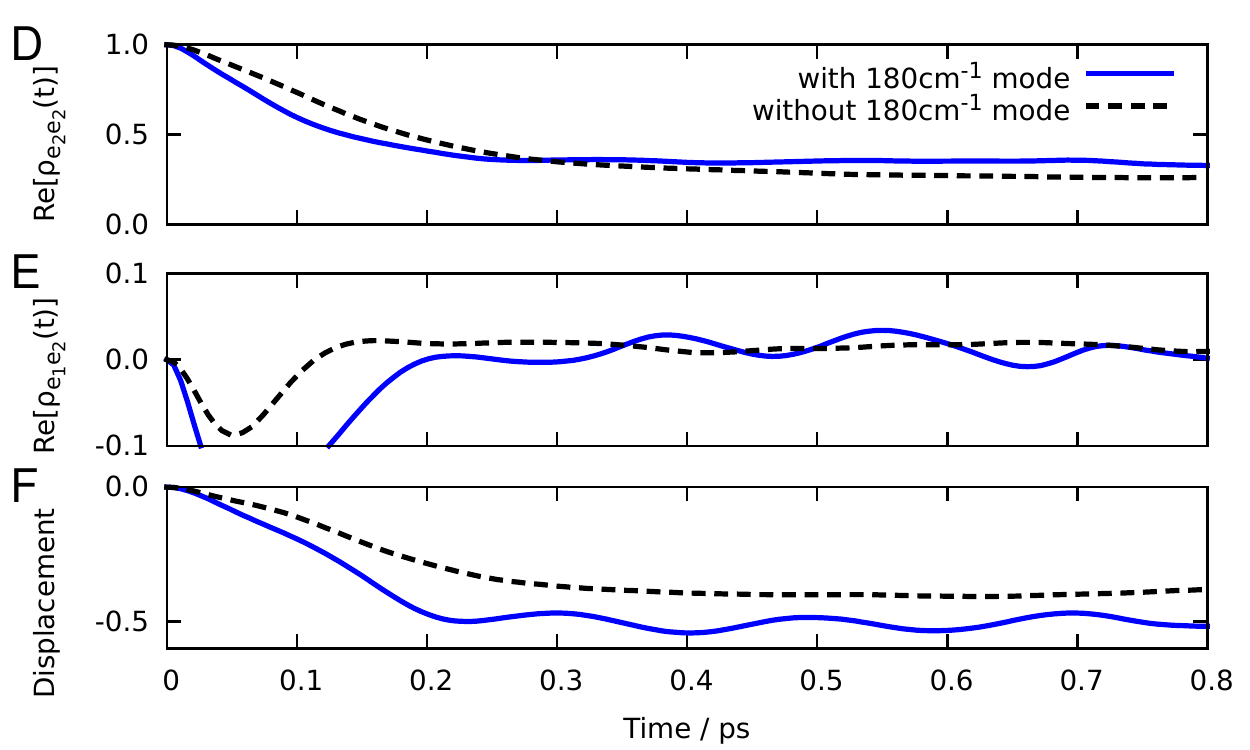}
\caption{ {\bf A}. Population dynamics, showing $\rho_{e_{2}e_{2}}(t)$ for electronic and mode parameters in the text. Initial state was $\rho(0)=|e_{2}\rangle\langle e_{2}|$ and environment was at $T=77$K. Weak, oscillatory revival of population in the $|e_{2}\rangle$ state between $0.3-0.6$ ps matches the revival dynamics of $\rho_{e_{2}g}$ in Fig. \protect\ref{exc1exc2} {\bf B}, indicating coherent population (and coherence) transfer. {\bf B}. spontaneous electronic coherences $Re[\rho_{e_{1}e_{2}}(t)]$ for same parameters and initial conditions.
{\bf C}. Average displacement $\langle X_{1}\rangle (t)$ for same parameters an initial conditions. {\bf D-F}. The same as {\bf A-C}, respectively, but with the environment initially at $T=277$K.}
\label{population}
\end{figure}

Figure \ref{exc1exc2}{\bf B} shows the absolute value of the ground-excited state coherences $\rho_{e_{n}g}(t)$ at $T=77$ K for initial
symmetric superpositions of states $|e_{n}\rangle$ and $|g\rangle$.
The thermal and quantum fluctuations of both the $180 \mathrm{cm}^{-1}$ and $37 \mathrm{cm}^{-1}$ modes induce
amplitude modulations of these coherences at the mode frequencies  \cite{caycedo2011nature}, leading to an effectively faster and oscillatory initial decay and then slow oscillations on top of the decaying amplitudes. The short effective time constants for all curves lie in the range $150-200$ fs, which are consistent with experimental results \cite{hayes2010dynamics}, and were faster than those found in the semiclassical approach.  The coherence
$\rho_{e_{2}g}(t)$ decays through population relaxation from $|e_{2}\rangle$ to $|e_{1}\rangle$,
however, on top of the periodic modulation caused by modes, additional weak recurrence features result from \emph{oscillatory population
transfer} induced by the $180 \mathrm{cm}^{-1}$ mode - see Fig. \ref{population}.  The coherence $\rho_{e_{1}g}(t)$
shows a fast, incomplete and non-markovian decay ($\approx 150$ fs) initially, and then an extremely slow decay due to the long-time absence of pure dephasing in the AR (super-Ohmic) spectral density and the long
($ \approx 2.4 $ ps) lifetime of the lowest energy excited state $|e_{1}\rangle$ at $T=77$ K.  Long $|e_1\rangle\langle g|$ coherence times ($10-100$ ps) have been seen in previous FMO photon echo measurements at low temperatures  ($T<50$ K) \cite{louvwe}. At $T=277$ K (Fig. \ref{exc1exc2} {\bf D}), both ground-excited coherences decay rapidly due to the fast population relaxation to thermal equilibrium and enhanced pure dephasing. Again, the additional oscillatory
decay in the presence of the resonant modes in the TEDOPA simulations leads to significantly
faster reduction of coherence (with time constant of $50$ fs) compared to the semiclassical
approach which neglects the quantum fluctuations of the environment.

\emph{Population oscillations and environmental variables}-- For $T=77$K, Fig.
\ref{population} {\bf A} shows the population remaining in $|e_{2}\rangle$ for the
initial state $\rho(0)=|e_{2}\rangle\langle e_{2}|$, which does not contain any
initial inter-exciton coherence and may be created by incoherent excitation alone.
The presence of the $180 \mathrm{cm}^{-1}$ mode induces population oscillations in
the population decay, which explain the similar oscillations seen in $\rho_{e_{1}g}(t)$
in Fig. \ref{exc1exc2} {\bf B}, and is also direct evidence of weak coherence transfer (not
shown).  Population oscillations have also recently been observed in FMO  \cite{panitchayangkoon2011direct}, and imply a
partially reversible energy exchange with the "environment" that transiently violates detailed balance.
Figure \ref{population} {\bf B}  shows the spontaneous generation of oscillatory coherences
over these population dynamics. Non-Markovian effects (finite reorganisation time) and the $37 \mathrm{cm}^{-1}$ mode are also seen to generate short-lived spontaneous coherences, however, these are transient and only
the $180 \mathrm{cm}^{-1}$ mode drives strong long-lasting coherence oscillations. No spontaneous coherence arises in the standard Bloch-Redfield description, showing the essential need for going beyond such theories - See Appendix \ref{nonmarkovsec}. In Figure \ref{population} {\bf C}, we make use of the complete environment information provided by TEDOPA
to show the expectation value of the collective environment coordinate $X_{1}$ \cite{prior10,chinchain10}.
Figure \ref{population} {\bf C} shows that the exciton-mode
interaction leads to long-lasting coherent environmental oscillations which drive exciton transitions, confirming the semiclassical picture and coherence-generating mechanism previously outlined. In
the absence of resonant modes, these strong oscillatory features vanish in the collective mode
and exciton populations. Figures \ref{population} {\bf D-F} show that, remarkably, the same non-equilibirum and spontaneously coherent dynamics are also present (though less strong) at $T=277$K.  

We finally observe that the spontaneous coherence in Fig. \ref{population} {\bf B} tends to a (small), non-oscillating, non-zero value at long times at $T=77$ K, indicating that the exciton-mode steady states generated by the dynamics are quantum superpositions of excitonic and vibrational degrees of freedom. This residual coherence vanishes at $T=277$ K. More striking examples of how mode interactions may alter relaxed electronic states is given in Appendix \ref{collini}, which sets out how discrete modes may provide insight into anomalous oscillations recently observed in conjugated polymers \cite{collini1,collini2}.

\emph{Single system vs ensemble--}
 In the linear absorption spectra of FMO ensembles, the lowest absorption
band - thought to consist of just a single excitonic transition, and possibly some vibrational sidebands \cite{Amerongen2000,Ratsep1999,ratsep2007electron,caycedo2011nature} - has an inhomogenous line
width of $80 \mathrm{cm}^{-1}$  FWHM \cite{Amerongen2000}. This implies an inhomogeneous
dephasing time of the $|e_{1}\rangle\langle g|$ coherences of $130$ fs, which is close to the experimental
measurements at $77$ K \cite{hayes2010dynamics}. In ensemble 2D spectroscopy experiments, this might completely mask the long (single complex) coherence time of $|e_{1}\rangle\langle g|$ at $T=77$ K in Fig. \ref{exc1exc2} {\bf B}, and -  moreover - if the inhomogeneous broadening of site energies on
each pigment is independent, then signals from \emph{inter-exciton} coherences such as
$|e_{1}\rangle\langle e_2|$
might also dephase on similar timescales due to the resulting width of inter-exciton energy differences \cite{van2000photosynthetic}.

The actual observation of long-lasting inter-exciton coherences in ensemble experiments thus requires further explanation, and could be supported by {\em correlated inhomogeneous broadening} that preserves energy \emph{differences}
between pigments in each complex, or by the dominant contribution of sub-ensembles with resonant exciton-vibration spectra to the oscillatory part of the experimental signals. Both ideas support an interpretation of the experimental quantum beats as arising from \emph{electronic coherence}, with coherence times extended by the mechanisms
presented in this article. In addition, the recent theory of  Christensson \textit{et al}. also demonstrates that hybridised exciton-vibrational states, which are generated spontaneously during our exact dynamical simulations, might indeed give robust (against inhomogeneous broadening) contributions to the oscillatory part of 2D spectra \cite{christensson2012origin} at $T=77$K. Experimental evidence for correlated inhomogeneous broadening has also been observed and discussed in the
FMO complex \cite{fidler2012two,caycedo2011nature}.

\emph{Conclusion --} We have demonstrated  that discrete resonant modes can induce electronic coherence
which is long-lasting and relevant in single complexes under a much wider range of initial
conditions than had been previously thought, including physiological temperatures and incoherent initial conditions. In focusing on purely electronic observables - which are the relevant degrees of freedom probed in actual EET experiments -  we have traced over the quantum fine structure of the hybrised exciton-mode
states, and have shown that an effective semiclassical motion of quasiresonant modes already provides
a qualitative and intuitive solution to the problem of how both long and short lasting electronic coherences may naturally appear in the PPCs. This heuristic model is subsequently corroborated by means of an exact numerical treatment that includes naturally the quantum fluctuations in the vibrational environment.
This insight illustrates a situation outside the standard open quantum system paradigm,
where parts of the environment (the {\em sharp} vibrational modes) are driven strongly out of equilibrium and counteract effects induced by the thermal background. The matching and coupling of underdamped vibrational frequencies to excitonic transitions presents a clear new design principle for building coherent exciton transport structures, and thus raises new possibilities within the context
of noise-assisted transport, such as  \emph{engineering} discrete bath structures for (coherent) energy transport -recently exemplified by the concept of the phonon antenna and recent studies of enhanced non-equilibrium energy harvesting efficiency away from detailed balance \cite{antennae,scully11}. From the point of view of fundamental physics, the experimental verification of the dynamics discussed in this work would provide additional ground for the consolidation of quantum biology
as a new, truly multidisciplinary research field.

\acknowledgements
This work was supported by the Alexander von Humboldt-Foundation, the EU STREP project PICC and
the EU Integrated Project Q-ESSENCE. AWC acknowledges support from the Winton Programme for the
Physics of Sustainability. JP was supported by Ministerio de Ciencia e Innovaci\'on Project No.
FIS2009-13483-C02-02 and the Fundaci\'on S\'eneca Project No. 11920/PI/09-j. We acknowledge the
BW-grid for computational resources. Aspects of this work have benefitted from discussions with
J. Almeida, A.~G. Dijkstra, D. Hayes, J. Caram, G.~S. Engel and R. van Grondelle.

\appendix
\section{Spectral functions, excitonic parameters and the finite-temperature TEDOPA method}
\subsection{Spectral function}\label{spectralfunction}
The spectral function used to characterise the environment acting on each site in the TEDOPA simulations is given by
\begin{eqnarray}
    J(\omega)&=& J_{0}(\omega) +\sum_{i=1}^{2}S_{i}\omega_{i}^{2}\delta(\omega-\omega_{i}),\label{spectral}\\
    J_{0}(\omega)&=&\frac{\lambda\,[1000\omega^{5}e^{-\sqrt{\frac{\omega}{\omega_{a}}}}+4.3\omega^{5} e^{-\sqrt{\frac{\omega}{\omega_{b}}}}]}{9!(1000\omega_{a}^{5}+4.3\omega_{b}^{5})}.
\end{eqnarray}
The bath parameters used throughout are $S_{1}=0.12, S_{2}=0.22, \omega_{a}=0.57 \mathrm{cm}^{-1}$
and $\omega_{b}=1.9 \mathrm{cm}^{-1}$. These parameters were extracted from experiments in
\cite{adolphs2006proteins,Ratsep1998}. Vibrational modes with similar frequencies have also
been experimentally observed in the FMO complex in \cite{Amerongen2000, ratsep2007electron}. The reorganisation energy used in our simulations is $\lambda=35 \mathrm{cm}^{-1}$, taken again from \cite{adolphs2006proteins}. The same spectral function is used for the semiclassical simulations, except the $37 \mathrm{cm}^{-1}$ mode is excluded.

\subsection{Simulation parameters and method}\label{dmrg}
The parameters of the dimer system used throughout are $\epsilon_{1}-\epsilon_{2}=130 \mathrm{cm}^{-1},
$ and $J= 53.5 \mathrm{cm}^{-1} $. Diagonalising the dimer
Hamiltonian yields two singly-excited exciton states $|e_{1,2}\rangle$ with an energy splitting
$E_{12}=E_{1}-E_{2}=170 \mathrm{cm}^{-1}$. Each constituent of the dimer is linearly coupled
to an environment of harmonic oscillators as in the well-known spin-boson model. This problem
is then transformed to a physically equivalent setting where each constituent of the dimer is coupled
to a linear chain of harmonic oscillators with nearest neighbor interaction \cite{prior10,chinchain10,dmrgbook}.
The initial state of the global system is always taken as $\rho_{total}=\rho(0)\otimes e^{-H_{B}/k_{b}T}$
i.e. a product of an arbitrary exciton state and the thermal state of the uncoupled environment at
temperature $T$. The initial thermal environment is prepared using imaginary time evolution
and the coupling between the system and thermal environment is turned on at time $t$ when the
initial thermal state of the environment has been prepared. Hence the problem is now the simulation
of the time evolution of this 1-D quantum system.
To this end the global state of the system and environment is then evolved using a finite-temperature
t-DMRG algorithm which retains information of the state of system and environment by combining
matrix product states (MPS) with a sophisticated way of implementing local updates to the system
yielding an efficient simulation of one-dimensional quantum systems (see Zwolak et al \cite{Zwolak} for
the description of imaginary time evolutions as well as real time evolutions in this framework).
The time dependence of arbitrary observables $O$ of the excitons or bath are then obtained from
$\langle O\rangle (t)=\mathrm{Tr}[\rho(t) O]$, where $\rho(t)$ is the total state of the system and
the bath.
For the present simulations the chains representing the environment are 49 sites long (preventing
recurrence effects on the relevant time scale) and are composed of harmonic oscillators truncated
to three levels. Using a bond dimension of $320$ for the MPS covers all essential physical effects
and restricts the required numerical resources to a reasonable limit.
The simulations have been carried out using a second order Suzuki-Trotter expansion with time steps
of $1/188$ and $1/136$ of the total time for the imaginary and real time evolution, respectively.
These parameters have been adjusted by observation of the norm during the simulation, as to keep
the incurred errors due to the truncation of the MPS and the Suzuki-Trotter expansion to a tolerable
value.
A detailed account - with several applications - of this new finite-temperature version of the
TEDOPA approach to open systems will be presented elsewhere.

\subsection{Exciton basis Hamiltonian and equations of motion}\label{hamex}
Rewriting the Hamiltonian of Eq. (\ref{ham}) in the exciton basis $|i\rangle=\sum_{n}C_{n}^{i}|e_{n}\rangle$, we obtain,
\begin{eqnarray}
    H &=& \sum_{n}E_{n}|e_{n}\rangle\langle e_{n}|+ \sum_{n, m}Q_{nm}|e_{n}\rangle\langle e_{m}|\nonumber\\
    &+& \sum_{ik}\omega_{k}a_{ik}^{\dagger}a_{ik}.\label{exham}
\end{eqnarray}
The new couplings to the bath are given by  $Q_{nm}=\sum_{i,k}g_{ik}C_{n}^{i}C_{m}^{i}(a_{ik}+a_{ik}^{\dagger})$.
The coefficients $C_{n}^{i}$ can be taken as real, so $Q_{nm}=Q_{mn}$. All bath modes are taken to be independent, i.e. $[a_{ik},a_{jl}^{\dagger}]=\delta_{ij}\delta_{kj}$, where the square bracket represents a commutator. In this basis, we see that the environment
now couples to both the energy level populations (diagonal coupling) and can induce transitions between exciton
states (off-diagonal coupling). Although transverse terms are important for the semiclassical driving of new exciton coherences by the modes, the diagonal terms are important in the initial excitation of the mode oscillations, especially when starting from an exciton population state.

\section{Semiclassical equations of motion}\label{semiclassics}
We have developed a theory in which the key process mediated by the discrete modes can be
interpreted as a quasi-coherent driving of the exciton system by the coherent displacement
of the modes. This suggests that a significant part of the interactions between the discrete
modes and the excitons could be accounted for in a semi-classical approximation. To derive
the semiclassical equations of motion which were used to investigate exciton dynamics in the FMO complex,
we treat the discrete modes as part of the system and account for the effects of the background
environment using a simple Bloch-Redfield approach. The effective system Hamiltonian is the same
as Eq. (\ref{exham}), but the sum over bath modes is restricted to just the two discrete modes
that appear in the spectral function of Eq. (\ref{spectral}). We then use the Heisenberg picture
to compute exact equations of motion for the mode displacements $X_{ik}(t)=\langle a_{ik}(t) +
a_{ik}^{\dagger}(t) \rangle$ and their momenta $P_{ik}(t)= \langle a_{ik}(t)-a_{ik}^{\dagger}(t)\rangle$,
where $A(t)=e^{iHt}Ae^{-iHt}$ denotes operators in the Heisenberg picture. The expectation value
refers to the operation $\langle A\rangle= \mathrm{Tr}[\rho_{total}(t)A]$. In this notation the
index $i$ refers to modes coupled to site $i$ and the $k$ index labels the different discrete
modes at each site i.e. the modes of frequency $\omega_{k}$ in Eq. (\ref{spectral}). As an example,
the equation of motion for the momentum operator is
\begin{eqnarray}
    \frac{d}{dt}(a_{ik}(t)-a_{ik}^{\dagger}(t))&=&i[H,(a_{ik}(t)-a_{ik}^{\dagger}(t))] \label{operatorequation}\\
    &=&-i\omega_{i} (a_{ik}(t)+a_{ik}^{\dagger}(t))\nonumber\\ &+&2i\sqrt{S_{k}}\omega_k\sum_{n}|C_{n}^{i}|^{2}|e_{n}\rangle
    \langle e_{n}|(t)\nonumber\\
    &+&2i\sqrt{S_{k}}\omega_k\sum_{n\neq m}C_{n}^{i}C_{m}^{i}|e_{m}\rangle \langle e_{n}|(t).
    \nonumber
\end{eqnarray}
Taking the expectation value of this equation of motion using the result
$\langle |e_{m}\rangle\langle e_{n}| (t)\rangle=\mathrm{Tr} [\rho_{total} (t)|e_{m}\rangle\langle e_{n}|]
=\rho_{e_{ n}e_{m}} (t)$, where $\rho_{e_{n}e_{m}}(t)$ are the matrix elements of the reduced
density matrix of the excitons, we arrive at the equation of motion for the expectation values
\begin{eqnarray}
    \frac{d P_{ik} (t)}{dt} &=& -i\omega_{i} X_{ik} (t)\nonumber\\ &+&2i\sqrt{S_{k}}\omega_k\sum_{n}|C_{n}^{i}|^{2}\rho_{e_{n}e_{n}}(t)\nonumber\\
    &+&2i\sqrt{S_{k}}\omega_k\sum_{n\neq m}C_{n}^{i}C_{m}^{i}\rho_{e_{n}e_{m}}(t)
    \label{momentum}
\end{eqnarray}
and analogously for the position operator
\begin{eqnarray}
    \frac{d X_{ik}(t)}{dt}&=&-i\omega_{k} P_{ik}(t) \label{position}.
\end{eqnarray}
The semiclassical approximation amounts to replacing the mode operators in Eq. (\ref{exham})
by these time-dependent c-number expectation values, creating a time-dependent Hamiltonian
$H_{X_{i}}(t)$ for the excitons. We evolve the reduced density matrix of the exciton system according to
\begin{eqnarray}
    \frac{d\rho_{s}(t)}{dt}&=&-i[H_{X_{i}}(t),\rho(t)]+ \mathcal{R}(t)[\rho_{s}(t)],\label{eqm}
\end{eqnarray}
where we have added a dissipative term $\mathcal{R}(t)$ that accounts for a background environment described by the smooth background part of the spectral function of Appendix \ref{spectralfunction}.
The Redfield relaxation tensor $\mathcal{R}(t)$ in Eq. (\ref{eqm}) is computed in the absence
of the discrete modes and is restricted to secular (energy-conserving) terms only. The full
equation and expressions for the relaxation and dephasing rates contained in $\mathcal{R}(t)$
are given explicitly in \cite{redfield,blum}. However, the time argument is included in $\mathcal{R}(t)$ to denote that some of the rates - corresponding to pure dephasing -  are non-standard, and are actually time-dependent. This will described in Appendix \ref{nonmarkovsec}.

In all simulations the initial mode displacements and momenta are taken drawn from a thermal distribution at temperature $T$. A phenomenological damping
rate $\gamma$ for the modes, which is assumed to be the same for both modes, is set to give
a typical damping time for vibrational motion of $1$ ps \cite{adolphs2006proteins,matsuzaki2000energy}. Thus Eq. (\ref{momentum}) is replaced by
\begin{eqnarray}
    \frac{d P_{ik} (t)}{dt} &=& -2\gamma P_{ik}(t)-i\omega_{i} X_{ik} (t)\nonumber\\ &+&2i\sqrt{S_{k}}\omega_k\sum_{n}|C_{n}^{i}|^{2}\rho_{e_{n}e_{n}}(t)\nonumber\\
    &+&2i\sqrt{S_{k}}\omega_k\sum_{n\neq m}C_{n}^{i}C_{m}^{i}\rho_{e_{n}e_{m}}(t).
    \label{momentumdamped}
\end{eqnarray}
For completeness, the equation of motion for the density matrix elements
$\rho_{e_{n}e_{m}}(t)=\langle e_{n}|\rho_{S}(t)|e_{m}\rangle$ are
\begin{eqnarray}
    \frac{d\rho_{e_{n}e_{m}}(t)}{dt} &=& -iE_{nm}\rho_{e_{n}e_{m}}(t) - i\sum_{j} Q_{nj}(t)
    \rho_{e_{j}e_{m}}(t)\nonumber\\
    &&\hspace*{-0.75cm}+ i\sum_{j} Q_{jm}\rho_{e_{n}e_{j}}(t) + \sum_{j,l}\mathcal{R}_{nmjk}\rho_{e_{j}e_{l}}(t),
    \label{exciton}
\end{eqnarray}
where $Q_{nm}(t)=\sum_{i,k}\sqrt{S_{k}}\omega_k C_{n}^{i}{C}_{m}^{i}X_{ik} (t)$. From Eq.
(\ref{exciton}) we see that the mode displacements lead to time-dependent couplings of
populations and coherences which generate the 'driving' of inter-exciton transitions
which are shown in the main text to enhance coherence lifetimes and generate new coherences. Oscillatory population
dynamics also appear through the coupling of populations to coherences mediated by the discrete modes. One can immediately see from the structure of these equations that discrete modes cannot drive $|e_{n}\rangle\langle g|$ coherences, and thus do not regenerate or support ground-excited coherence. However, these equations do predict $\rho_{e_{n}g}\rightarrow \rho_{e_{m}g}$ coherence transfers due to
coherent motion of the discrete modes which lead to rapid suppression of coherence except for the lowest exciton. For each set condition of initial mode conditions drawn from the thermal distribution,
the equations of motion for $X_{i}(t),P_{i}(t)$ eqs. \ref{position},\ref{momentumdamped}
and eq. \ref{exciton} are integrated together, and these trajectories are averaged over to obtain the results of Fig. \ref{semis}.

We finally note that in this simple description, quantum fluctuations of the
discrete modes are neglected, as are the renormalisation effects of the background environment.
This last approximation is appropriate for the relatively weak coupling parameters used for the background
spectral density. All of these effects are captured in the TEDOPA simulations. The main qualitative differences were observed in the $|e_{n}\rangle\langle g|$ coherences, where amplitude modulations due to quantum fluctuations of the discrete modes could not be reproduced by the semiclassical approach, leading to an overestimate of coherence times.

\begin{figure}
\includegraphics[width=8.5cm]{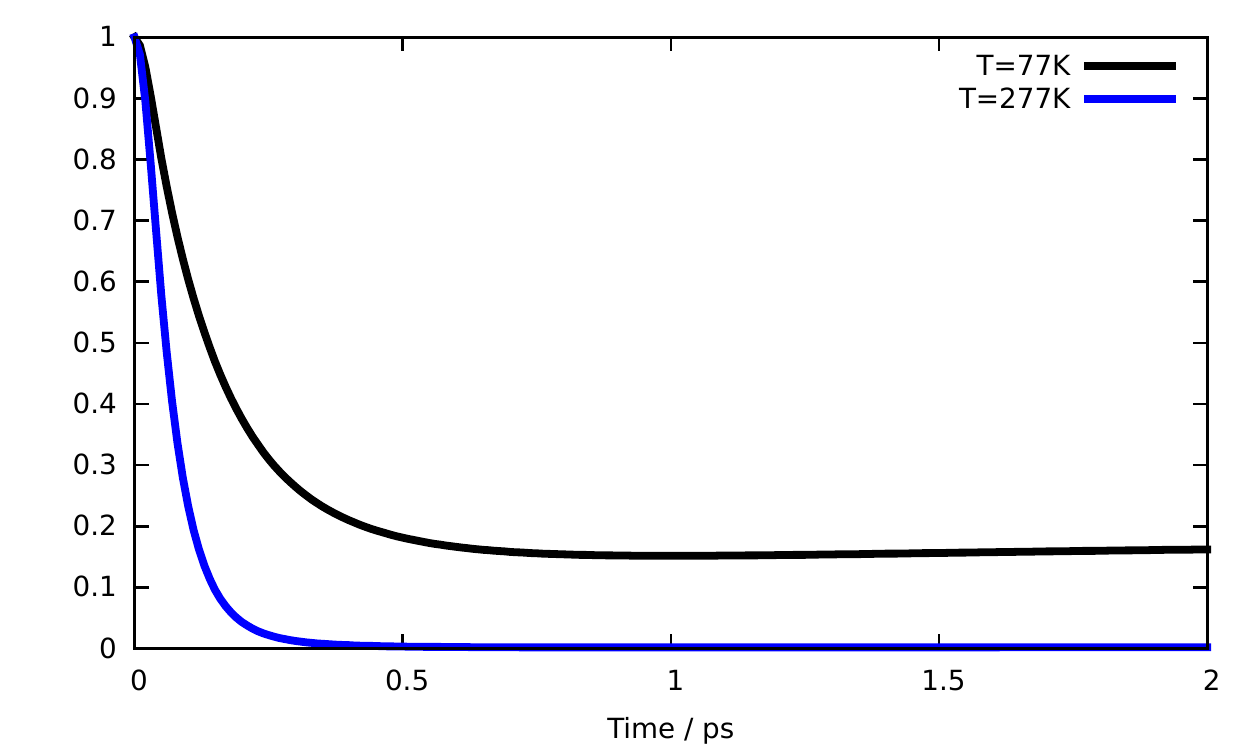}
\caption{$\rho_{12}(t)=\rho(0)e^{-\int_{0}^{t}\Gamma_{pd}(t')dt'}$. Initial condition is $\rho(0)=1$ and the transient, non-Markovian pure dephasing rate is computed using Eq. (\ref{nonmarkov2}) for the spectral density given in \ref{spectralfunction}. Results are shown at $T=77$K (black line) and $T=277$K (blue line). The vanishing of the dephasing rate at long times leads to the survival of a finite value of coherence at long times.    }
\label{nonmarkovplot}
\end{figure}

\begin{figure}
\includegraphics[width=8.5cm]{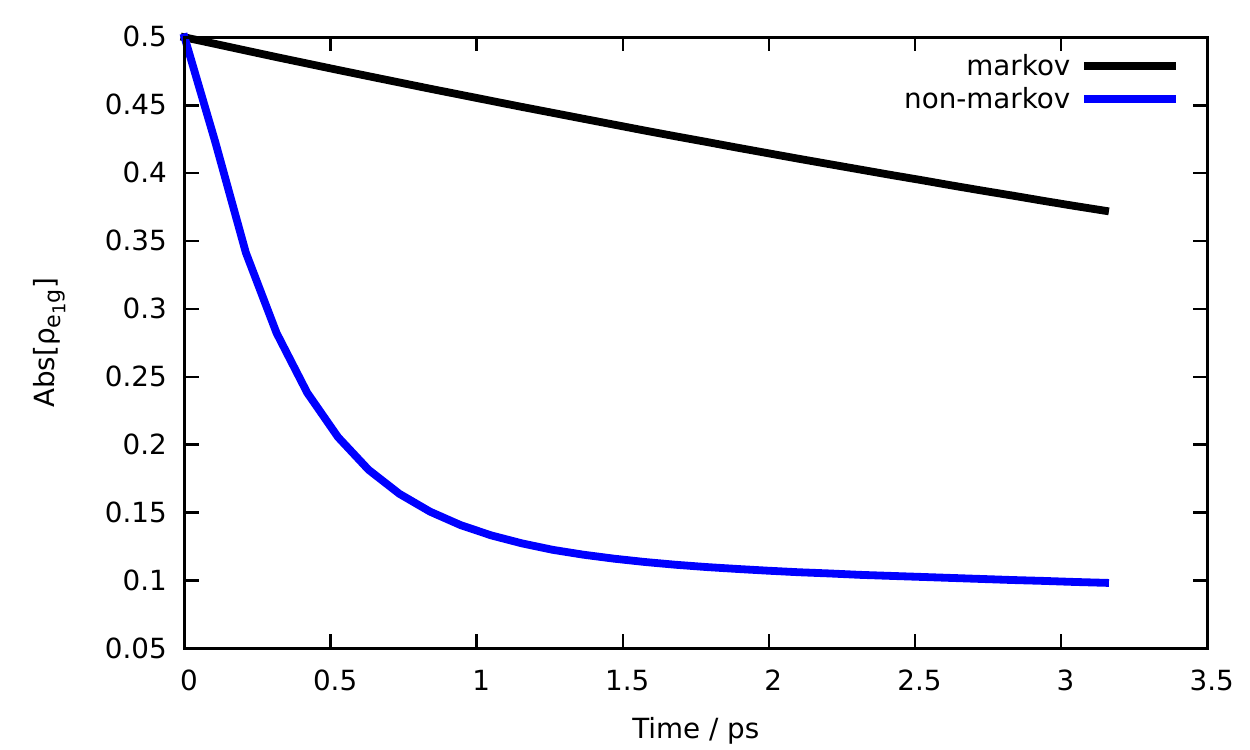}
\caption{$\mathrm{Abs}[\rho_{e1g}(t)]$, for the standard, secular Bloch-Redfield theory (Black line) and for the generalised theory which includes a time-dependent pure dephasing rate (Blue line). Initial condition is $2\rho(0)=|e_{i}\rangle\langle e_{i}|+
|g\rangle\langle g|+|e_{i}\rangle\langle g|+|g\rangle\langle e_{i}|$ and the bath is at $T=77$K. Calculations are for the full FMO system described in the main text.}
\label{e1gnonmarkov}
\end{figure}
\subsection{Non-Markovian Pure dephasing rates for  super-Ohmic spectral densities}\label{nonmarkovsec}
In standard, secular Bloch-Redfield theory, the Markovian pure dephasing rate $\Gamma_{pd}$ (which causes decoherence of energy eigenstate superpositions in the absence of energy relaxation) at temperature $T$ for a spectral density $J(\omega)$ is (neglecting matrix elements) proportional to

\begin{eqnarray}
\Gamma_{pd}&\propto&\lim_{\epsilon \rightarrow 0}\int_{0}^{\infty} dt\int_{0}^{\infty} d\omega J(\omega)\coth\left(\frac{\beta\omega}{2}\right)\cos\left( (\omega-\epsilon)t\right),\nonumber\\
&\propto& \lim_{\epsilon \rightarrow
0}\pi\int_{0}^{\infty} d\omega J(\omega)\coth\left(\frac{\beta\omega}{2}\right)\delta(\omega-\epsilon),\nonumber\\
&\propto&\lim_{\epsilon \rightarrow
0}\left[\pi J(\epsilon)\coth\left(\frac{\beta\epsilon}{2}\right)\right],\nonumber\\
&\propto&\lim_{\epsilon \rightarrow
0}\left[2\frac{\pi}{\beta\epsilon} J(\epsilon)\right],\label{puredephasing}
\end{eqnarray}
where $\beta=(k_{b}T)^{-1}$. The pure dephasing rate is thus controlled by the behaviour of spectral function approaching zero frequency \cite{brumer,kramer}, as one would expect for a process that involves no energy dissipation. Typical spectral densities have power-law dependence at low frequency $J(\omega)\propto \omega^{s}$, with $s$ some exponent. For super Ohmic damping ($s>1$) the limit in Eq. (\ref{puredephasing}) is zero, and thus pure dephasing is apparently absent for this class of spectral functions \cite{kramer}. This can considerably weaken the temperature dependence of dephasing, which can now only occur through energy relaxation in Bloch-Redfield theory \cite{brumer}. The AR spectral function of \ref{spectralfunction} is in this class. The commonly encountered Ohmic spectral density ($s=1$) leads to a finite limit and a pure dephasing rate proportional to temperature \cite{brumer}. The relevance of this observation has recently been discussed in the context of PPC dynamics \cite{brumer,kramer}, and the absence of pure dephasing in super-Ohmic spectral densities has been put forward as a means of prolonging excitonic coherence whilst maintaining efficient energy transport \cite{kramer}.

However, the pure dephasing \emph{rate} goes to zero in the full Markov approximation due to the fact that the upper limit in the integral over time in Eq. (\ref{puredephasing}) is taken to infinity, leading to the delta function that picks out the zero frequency component. However, a more accurate treatment would take the upper limit to be $t$ (finite) so that the (now time-dependent) pure dephasing rate also remains finite (it samples a \emph{range} of finite frequencies proportional to $1/t$) at short times, and, at least at those early times, may contribute significantly to the dephasing of coherences. To take account of this, we therefore introduce a time-dependent pure dephasing rate $\Gamma_{pd}(t)$ in place of the vanishing Markovian rates in our Redfield tensor proportional to

\begin{equation}
\Gamma_{pd}(t)\propto\int_{0}^{t} dt\int_{0}^{\infty} d\omega J(\omega)\coth\left(\frac{\beta\omega}{2}\right)\cos(\omega t).
\label{nonmarkov2}
\end{equation}

The key point is that while $\Gamma_{pd}(t)\rightarrow 0$ as $t\rightarrow \infty$, the evolution of a coherence $\rho_{12} (t)$ obeying the equation of motion $\dot{\rho}_{12}(t)=-\Gamma_{pd}(t)\rho_{12}(t)$ is \cite{metrology}
\begin{equation}
\rho_{12}(t)=\rho_{12}(0)e^{-\int_{0}^{t}\Gamma_{pd}(t')dt'},
\end{equation}
and the exponential suppression resulting from the integral over the transient time-dependent rate can be non-negligible. Figure \ref{nonmarkovplot} illustrates this, showing this suppression factor for the AR background spectral density at $T=77K$ and $T=277$K, taking $\Gamma_{pd}(t)$ to be equal to the RHS of Eq. (\ref{nonmarkov2}). Theses curves shows a fast initial suppression which then stops and plateaus as the pure dephasing rate vanishes. The `residual' coherence is suppressed at higher temperature, in line with expectations based on exactly solvable, non-Markovian pure dephasing models which also predict the residual coherence for super-Ohmic spectral densities \cite{metrology}. Despite the long-time absence of pure-dephasing, Fig. \ref{nonmarkovplot} clearly demonstrates the importance of accounting for the early time suppression, and naive applications of standard Bloch-Redfield theory for super-Ohmic spectral densities are likely to over-estimate effective coherence times without this correction.  

\begin{figure}
\includegraphics[width=8.5cm]{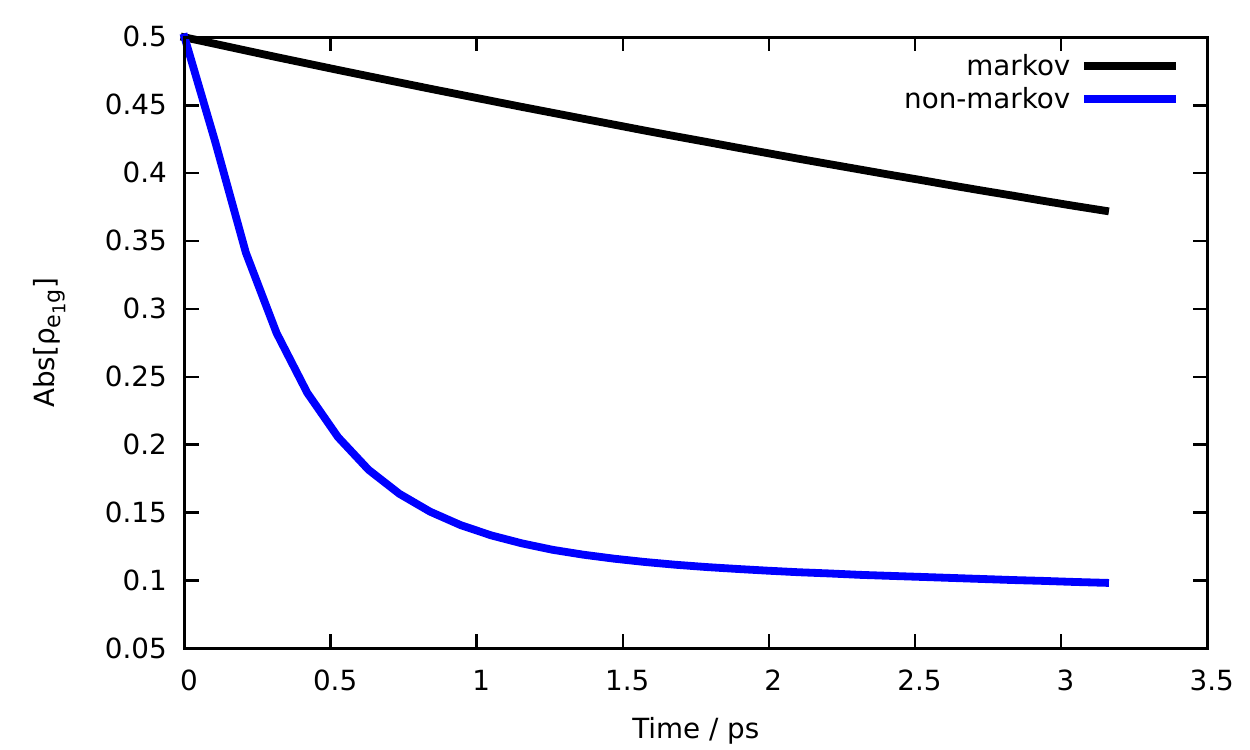}
\caption{$\mathrm{Abs}[\rho_{e1g}(t)]$, for the standard, secular Bloch-Redfield theory (Black line) and for the generalised theory which includes a time-dependent pure dephasing rate (Blue line). Initial condition is $2\rho(0)=|e_{i}\rangle\langle e_{i}|+
|g\rangle\langle g|+|e_{i}\rangle\langle g|+|g\rangle\langle e_{i}|$ and the bath is at $T=77$K. Calculations are for the full FMO system described in the main text.}
\label{e1gnonmarkova}
\end{figure}

We illustrate this further with Fig. \ref{e1gnonmarkova}, which shows $Abs[\rho_{e_{1}g}(t)$ for the full FMO complex computed with and without the super-Ohmic time-dependent pure dephasing. The Bloch-Redfield lifetime of the (lowest exciton energy) $|e_{1}\rangle$ state at $T=77K$ is $2.4$ ps, implying a very long  dephasing time of $4.8$ ps in the absence of pure dephasing. This is completely inconsistent with the TEDOPA results in Fig. \ref{exc1exc2}, while the corrected non-Markovian dephasing result actually gives much better agreement w.r.t overall dephasing rates and the long-time value of the coherence. One sees that the dephasing is dominated by the early-time non-Markovian correction, with the much weaker dephasing arising from the relaxation of the $|e_{1}\rangle$ state leading to a slow decay of the residual coherence plateau.

We note, however, that the inclusion of a time-dependent, non-markovian pure dephasing rate in our simple time-local master equation is an approximation, and for some spectral densities the time-local master equation may be invalided.  Fortunately, this important, transient pure dephasing effect is handled exactly by TEDOPA, and the results of this section are provided to show the necessity of using advanced techniques for the open-quantum dynamics of PPCs, as well as the improvements to numerically cheaper techniques that insight from TEDOPA can generate.

\section{ Anomalous anisotropy decay in organic polymers and spontaneous generation of coherences}
\label{collini}
\begin{figure}
\includegraphics[width=8cm]{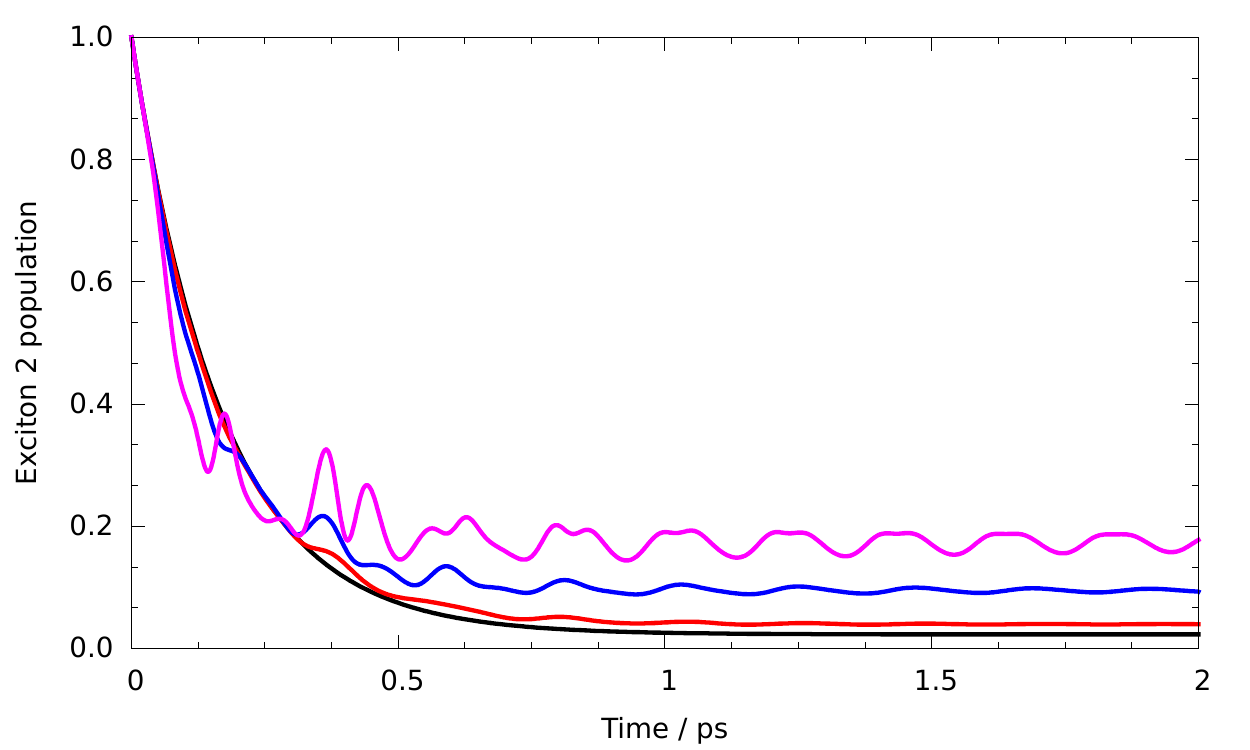}
\caption{Exciton $2$ population for initial condition
$\rho_{s}(0)=|e_{2}\rangle\langle e_{2}|$ using the semiclassical equations at $T=77$K,
$t_{vib}=1$ ps and $S=0.00,0.22, 0.50, 1.00$ (black, red, blue, magenta, respectively). Horizontal axis denotes time in ps.}
\label{semirabi}
\end{figure}

In a recent paper Collini and Scholes demonstrated room-temperature inter-exciton coherence in
conjugated polymers in chloroform solution \cite{collini1,collini2}. They present some anomalous
anisotropy decay data in which long-lasting ($\ge 1$ ps) coherent oscillations and bi-exponential
population transfer in the anisotropy decay were observed \cite{collini2}. These oscillations had
frequencies consistent with vibrational frequencies in the system, but were also rather similar
to exciton energy differences as well. The data showed conflicting features (lack of coherence time
dependence, long-lasting oscillations) which meant that it was not possible to assign these features
unambiguously to either electronic or vibrational coherences. The model and mechanisms presented
here may provide some insight into the these data, as they show how the generation of spontaneous
coherences by the mode-exciton interaction may induce population oscillations on long time scales
even when the exciton system is populated without exciting vibrational coherence. Figure \ref{semirabi}
shows some illustrative population dynamics obtained using the semiclassical approach of section
\ref{semiclassics} for a dimer with $\epsilon_{1}-\epsilon_{2}=100\mathrm{cm}^{-1}, J_{12}=86\mathrm{cm}^{-1}$, each site of which
is coupled to a discrete mode with $\omega_{i}=180 \mathrm{cm}^{-1}$ and a variable Huang-Rhys
factor $S$. The background bath is at zero temperature. Two key features which emerge in the presence of the mode interaction is that the population
decay develops a number of new timescales (relative to the case $S_{i}=0.0$). The mode leads to a
multi-exponential population decay (initially faster when modes are present) on which complex coherent
oscillations resulting from the mode interactions are superimposed. These coherent oscillations
persist for at least the dephasing time of the modes ($1$ ps in these simulations), and are stronger
and persist longer at stronger coupling.

\begin{figure}
\includegraphics[width=8cm]{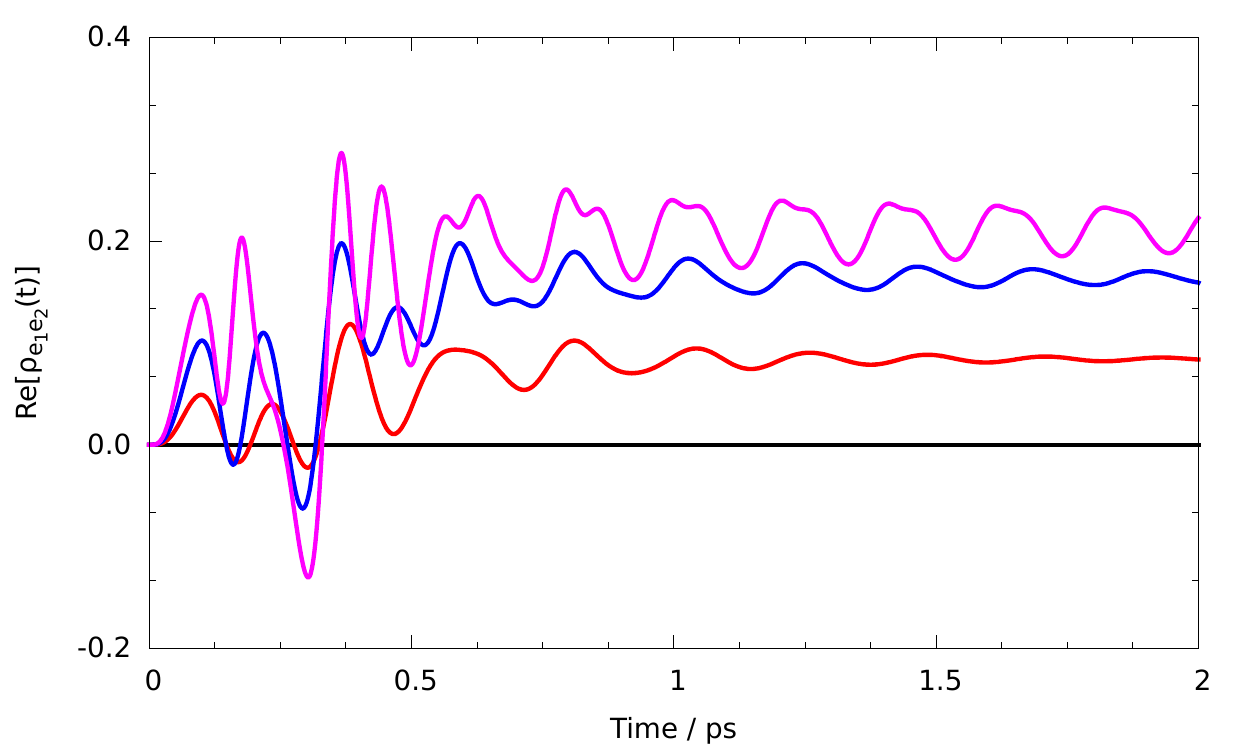}
\caption{Spontaneously generated coherences $\mathrm{Re}[\rho_{e_{1}e_{2}}(t)]$ for initial condition
$\rho_{s}(0)=|e_{1}\rangle\langle e_{1}|$ using the semiclassical equations at $T=77$K, $t_{vib}=1$ ps
and $S=0.00, 0.22, 0.50, 1.00$ (black, red, blue, magenta curves respectively). Horizontal axis denotes time in ps.}
\label{semispontcoh}
\end{figure}

Figure \ref{semispontcoh} shows the spontaneous generation of the electronic coherence
by the discrete modes for the same parameters and initial conditions as in Fig. \ref{semirabi}.
This provides direct evidence that the population oscillations shown in Fig. \ref{semirabi}
are the result of inter-exciton (electronic) coherences induced by the driving of the discrete modes.
The magnitude, oscillation lifetime and final value of these spontaneous coherences increase
with coupling strength, and show interesting dynamical oscillations at very strong coupling,
where weak signs of non-rotating wave harmonics can be seen in the driven coherences. The
final values show that the discrete modes change the nature of the equilibrium exciton states and both exciton states are significantly hybridised by the modes at strong coupling. Note that these calculation are intended to illustrate the physics which might explain the data of Refs. \cite{collini1,collini2}, but the parameters we have chosen are fairly arbitrary and a much more detailed analysis of the physical and experimental set up is required to unambiguously assign  a mechanism to the observed behaviour.

\end{document}